\documentclass[%
reprint,
superscriptaddress,
bibnotes,
amsmath,amssymb,
prb,
floatfix,
]{revtex4-2}

\usepackage{graphicx}
\usepackage{dcolumn}
\usepackage{bm}
\usepackage{hyperref}

\usepackage{xcolor}

\newcommand{\dif}{\text{d}}

\newcommand{\dert}[3]{\frac{\text{d}^#3 #1}{\text{d} #2^#3}}

\newcommand{\hn}{\mathcal{H}\text{N}}
\newcommand{\HN}{Struve-Neumann }
\newcommand{\RK}{Rytova-Keldysh }
\newcommand{\qcRK}{quantum-confined Rytova-Keldysh }
\renewcommand{\vec}[1]{\boldsymbol{#1}}

\begin{document}

\preprint{}

\title{Effects of Two-Dimensional Material Thickness and Surrounding Dielectric Medium on Coulomb Interactions and Excitons}

\author{F. Garc\'ia Fl\'orez}
\email{f.garciaflorez@uu.nl}
\affiliation{Institute for Theoretical Physics and Center for Extreme Matter and Emergent Phenomena, Utrecht University, Princentonplein 5, 3584 CC Utrecht, The Netherlands}

\author{Laurens D. A. Siebbeles}
\email{l.d.a.siebbeles@tudelft.nl}
\affiliation{Optoelectronic Materials Section, Department of Chemical Engineering, Delft University of Technology, Van der Maasweg 9, 2629 HZ, Delft}%

\author{H. T. C. Stoof}
\email{h.t.c.stoof@uu.nl}
\affiliation{Institute for Theoretical Physics and Center for Extreme Matter and Emergent Phenomena, Utrecht University, Princentonplein 5, 3584 CC Utrecht, The Netherlands}

\date{\today}

\begin{abstract}
    We examine the impact of quantum confinement on the interaction potential between two charges in two-dimensional semiconductor nanosheets in solution.
    The resulting effective potential depends on two length scales, namely the thickness $d$ and an emergent length scale $d^* \equiv \epsilon d / \epsilon_{\text{sol}}$, where $\epsilon$ is the permittivity of the nanosheet and $\epsilon_{\text{sol}}$ is the permittivity of the solvent.
    In particular, quantum confinement, and not electrostatics, is responsible for the logarithmic behavior of the effective potential for separations smaller than $d$, instead of the one-over-distance bulk Coulomb interaction.
    Finally, we corroborate that the exciton binding energy also depends on the two-dimensional exciton Bohr radius $a_0$ in addition to the length scales $d$ and $d^*$ and analyze the consequences of this dependence.
\end{abstract}

\maketitle


\section{\label{sec:introduction}Introduction}

Research into two-dimensional materials has increased in recent years, driven in particular by prospects for their use in new state-of-the-art optoelectronic devices that convert light into electric current and vice versa \cite{goryca2019,mak2010,wang2012,bernardi2013,jariwala2014,bie2017,manser2016,ye2015,xia2014,kagan2016,bonaccorso2010,grim2014,yang2017,pelton2018,tomar2019}.
However, understanding these devices and making them more efficient requires a firm foundation as a starting point.
In particular, two-dimensional semiconductor nanosheets in solution are studied in pump-probe experiments, in which electrons and holes are created using a pump laser and their nature and properties are subsequently characterized with a probe laser measuring the complex conductivity \cite{tomar2019}.\ Since the presence of excitons may make-or-break a particular application for optoelectronic devices, a refined understanding of the properties of excitons, such as their mass, average size, and binding energy, is advantageous.
The most important aspect that greatly affects the dynamics of excitons is the attractive interaction potential between electrons and holes that allows the bound state to form.
Electron-hole interactions in two-dimensional materials have therefore been considered for several decades, resulting in the \RK potential that has been extensively used, and extended, in the literature thus far \cite{rytova2018,keldysh1979,levine1982,rohlfing1998,cudazzo2011,schmitt-rink1985a,latini2015,trolle2017}.
Our goal in this paper is to better understand the consequences of introducing quantum confinement into the electron-hole interaction potential, specifically its role regarding the short-distance logarithmic behavior expected for a purely two-dimensional Coulomb potential, and ultimately also on the exciton properties.
Our approach in particular discusses the importance of the three length scales involved in the exciton problem, i.e., the thickness of the nanosheet $d$, the emergent length scale $d^* \equiv \epsilon d / \epsilon_{\text{sol}}$ from electrostatics that typically is much larger than the thickness $d$ as the permittivity $\epsilon$ of the nanosheet is much larger than the permittivity $\epsilon_{\text{sol}}$ of the solvent, and the two-dimensional exciton Bohr radius $a_0$ that is introduced by quantum mechanics due to the relative kinetic energy of the electron-hole pair.

\begin{figure}[h!]
    \begin{center}
        \includegraphics[width=\linewidth]{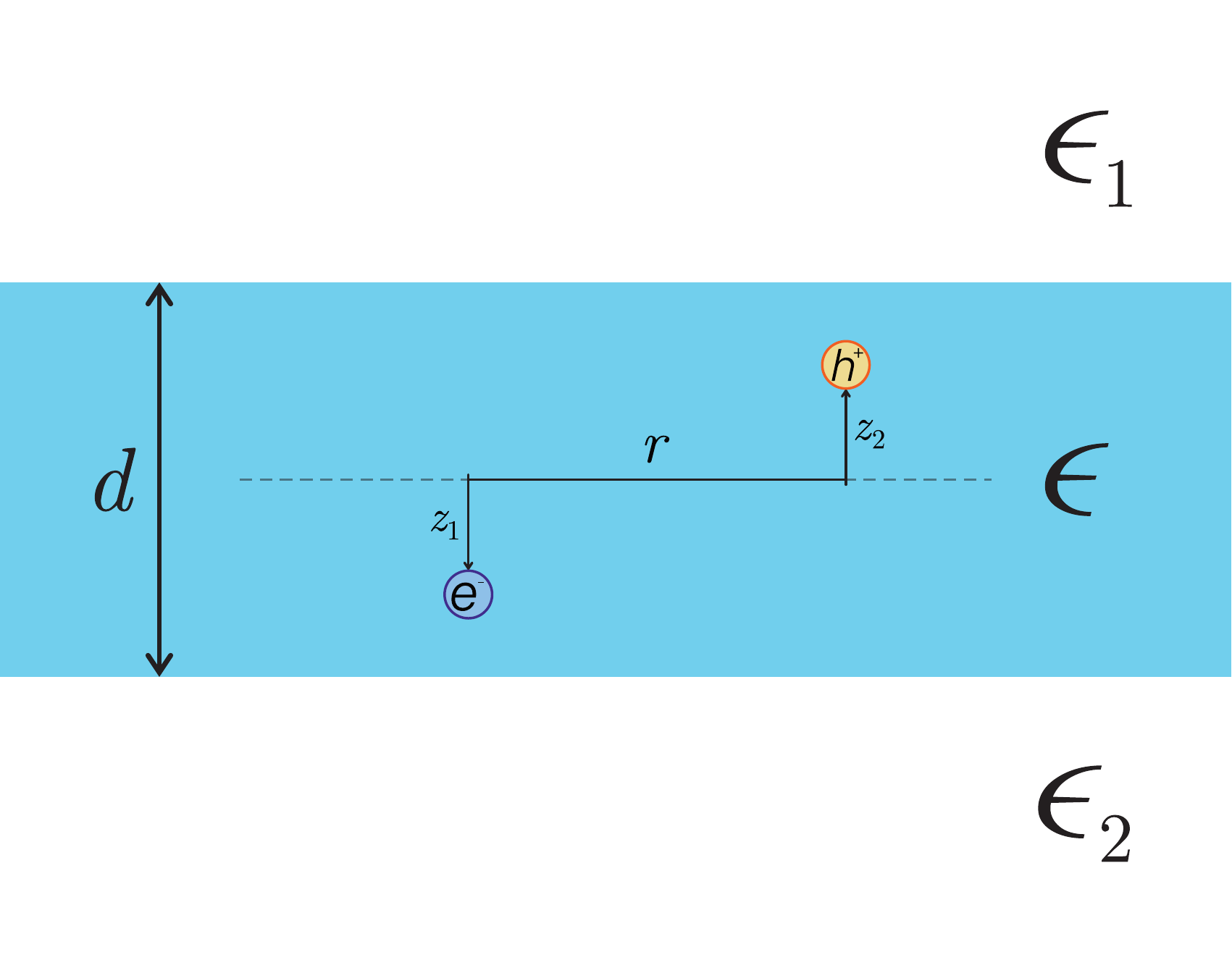}
        \caption{
            Diagram of the system under consideration.
            A nanosheet of thickness $d$ and permittivity $\epsilon$ is surrounded by an environment above and below with permittivities $\epsilon_1$ and $\epsilon_2$ respectively.
            The \RK potential is the solution of the electrostatics problem for the interaction potential between the two charges separated by an in-plane distance $r$, and displaced along the perpendicular axis by a distance $z_1$ and $z_2$, respectively.
        }
        \label{fig:scheme}
    \end{center}
\end{figure}

Initially presented in Refs.\ \cite{rytova2018,keldysh1979}, the electrostatic \RK potential indeed incorporates both length scales $d$ and $d^*$, as we will see explicitly in a moment.
More specifically, the \RK potential is the solution of the electrostatics problem that describes the electron-hole interactions in a nanosheet of permittivity $\epsilon$ and thickness $d$, surrounded by an environment of permittivities $\epsilon_1$ and $\epsilon_2$.
Figure \ref{fig:scheme} shows an artist impression of this configuration.
An analytic expression that approximates the \RK potential and that is widely used to describe interactions in two-dimensional materials, is obtained in the large distance limit $r \gg d$ and $\epsilon \gg \epsilon_{1,2}$.
This analytic expression is

\begin{align}
    \begin{split}
        V^{\hn}(r) &= - \frac{e^2}{4 \pi \epsilon} \\
                                   &\times \frac{\pi}{d} \left[ \mathcal{H}_0\left(\frac{\epsilon_1+\epsilon_2}{\epsilon} ~ \frac{r}{d} \right) - N_0 \left( \frac{\epsilon_1+\epsilon_2}{\epsilon} ~ \frac{r}{d} \right) \right]
        ~~ ,
    \end{split}
\end{align}

\noindent that we denote as the \HN (``$\hn$'') potential.
Here $\mathcal{H}_{\nu}(x)$ is the Struve function and $N_{\nu}(x)$ is the Neumann function, also known as the Bessel function of the second kind.
In order to expose its universal properties, we make the potential dimensionless by dividing by the Coulomb energy $E^C \equiv e^2 / 4 \pi \epsilon d$, which results in the dimensionless potential

\begin{equation}
    \label{eq:r_hn_pot}
        V^{\hn}(r) = - \pi \left[ \mathcal{H}_0\left(\frac{2r}{d^*} \right) - N_0 \left(\frac{2r}{d^*} \right) \right]
        ~~ ,
\end{equation}

\noindent where we have defined $d^* \equiv 2\epsilon d / (\epsilon_1 + \epsilon_2)$, thus explicitly showing that the dimensionless \HN potential only depends on the ratio $r / d^*$.
In other words, the dimensionless \HN potential does not depend on $d$ and $d^*$ independently but only on the latter, consistent with the limit $r \gg d$.
Note that from now on every energy is made dimensionless in the same manner.
For the purpose of simplicity in the discussion of our results we consider only the case of nanosheets in solution with $\epsilon_1 = \epsilon_2 \equiv \epsilon_{\text{sol}} \leq \epsilon$, while equations are given for the more general situation $\epsilon_1 \neq \epsilon_2$.

Our paper is organized as follows.
Section \ref{sec:electrostatics} revisits the properties of the \RK potential, that is, both its small and large-distance behavior.
Section \ref{sec:eff_pot} presents a derivation of the potential that incorporates the effect of quantum confinement, and it is applied to both the Coulomb potential, which is valid for the special case $\epsilon_{\text{sol}} = \epsilon$, and to the full \RK potential for which $\epsilon_{\text{sol}} \neq \epsilon$.
Finally, Sec. \ref{sec:excitons} analyzes the exciton binding energy computed using each of the potentials presented in the previous sections, and Sec. \ref{sec:discussion} concludes with a discussion about our findings.

\section{\label{sec:electrostatics}Electrostatics}

The full \RK potential, after dividing by the Coulomb energy $E^C$, is given in momentum space by \cite{keldysh1979,rytova2018}

\begin{align} \label{eq:k_ke_pot}
    \begin{split}
        V^{\text{RK}}&(k, z_1, z_2) = - \frac{2 \pi d^2}{k d} \\
        &\times \frac{2\cosh\left[k \left(\frac{d}{2} - z_1\right) + \eta_2\right] \cosh\left[k \left(\frac{d}{2} + z_2\right) + \eta_1\right]}{\sinh\left(k d + \eta_1 + \eta_2 \right)} ~~,
    \end{split}
\end{align}

\noindent where

\begin{equation}
    \eta_a \equiv \frac{1}{2} \ln\left(\frac{\epsilon + \epsilon_a}{\epsilon - \epsilon_a}\right)
    ~~ , ~~~ \text{ for } a \in \{1,2\}
    ~~.
\end{equation}

\noindent The coordinates $z_1$ and $z_2$ correspond to the position of the two charges along the axis perpendicular to the plane as shown in Fig. \ref{fig:scheme}, and are both located in the interval $[-d/2, d/2]$.
Furthermore, we define what is usually called the \RK potential as $V^{\text{RK}}(k) \equiv V^{\text{RK}}(k, 0, 0)$.

\begin{figure}[h!]
    \begin{center}
        \includegraphics[width=\linewidth]{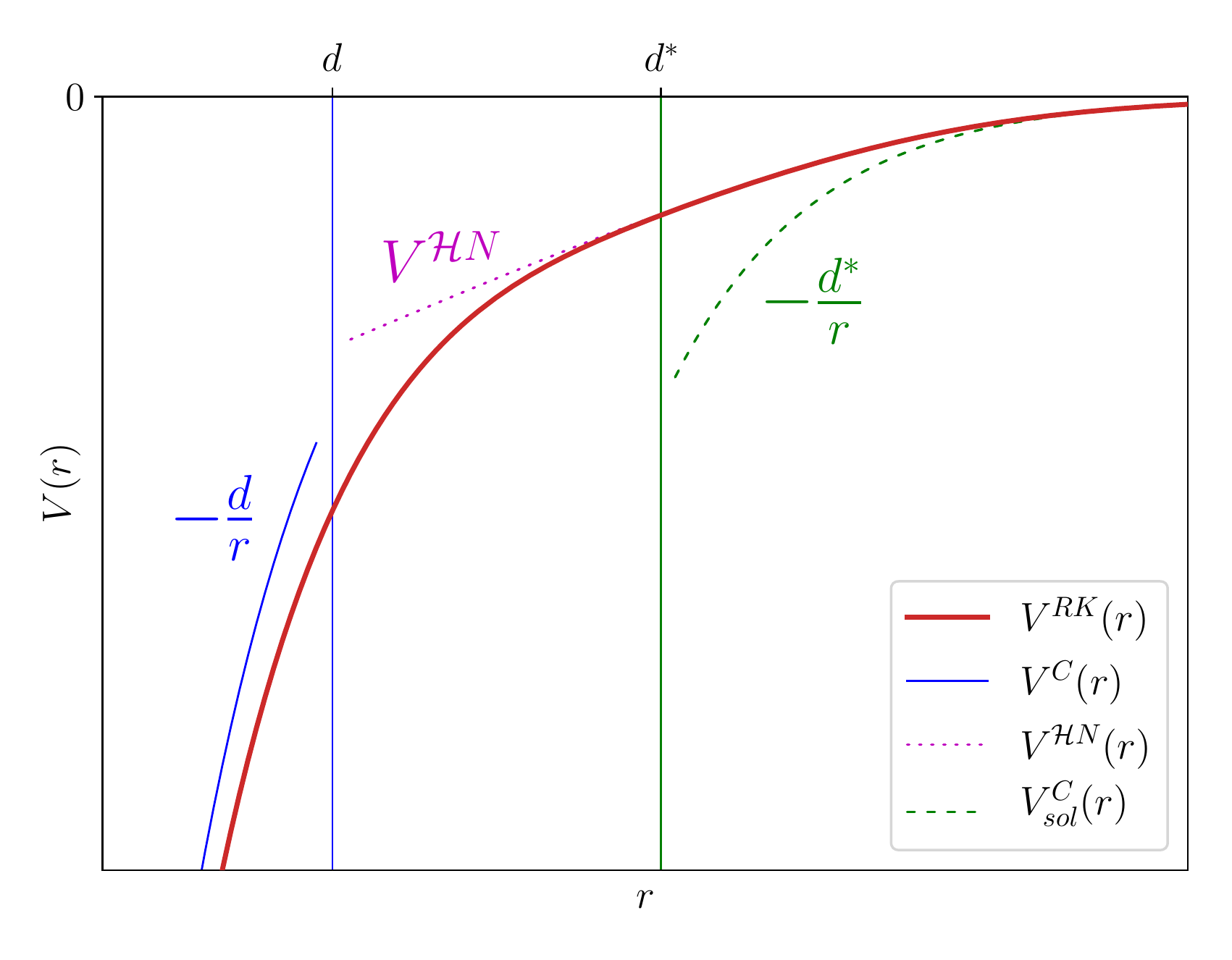}
        \caption{
            The \RK potential compared to the three approximations at small, intermediate, and large distances.
            In the regime $r \gg d$ the \RK potential is approximated by the \HN potential, given in Eq. (\ref{eq:r_hn_pot}).
            In the regime $r \gg d^*$ the \HN potential further reduces to the Coulomb potential of the solvent, given in Eq. (\ref{eq:r_ke_pot_long}).
            In the regime $r \ll d$ the Coulomb potential of the semiconductor material is recovered, given in Eq. (\ref{eq:r_ke_pot_short})
        }
        \label{fig:pot_scheme}
    \end{center}
\end{figure}

To study the \RK potential in real space Eq. (\ref{eq:k_ke_pot}) must be Fourier transformed numerically for $z_1 = z_2 = 0$, which Fig. \ref{fig:pot_scheme} shows as a solid red line.
The result contains three different regions separated by the lengths $d$ and $d^*$, i.e., $r \ll d$, $r \gg d$, and at even larger distances $r \gg d^*$, each resulting in a different approximation of the \RK potential.
In the regime $r \gg d^*$ the \RK potential reduces to the Coulomb potential with the permittivity of the solvent, that in our units is expressed as

\begin{equation} \label{eq:r_ke_pot_long}
    V^{\text{RK}}(r \gg d^*) \simeq - \frac{d^*}{r} \equiv V_{\text{sol}}^{\text{C}}(r)
    ~~ .
\end{equation}

\noindent Notice that even though $V^{\text{C}}_{\text{sol}}(r)$ seems to depend on the thickness $d$ and material permittivity $\epsilon$, it is only but a byproduct of scaling by $E^C$ --- in S.I. units this potential is a function of $\epsilon_{\text{sol}}$ alone, that is,

\begin{equation}
    V_{\text{sol}}^{\text{C}}(r) = - \frac{e^2}{4 \pi \epsilon_{\text{sol}}} \frac{1}{r}
    \hspace{1cm}
    \text{ (in S.I. units)}
    .
\end{equation}

\noindent The \HN potential approximates the \RK potential for distances $r \gg d$, as Fig. \ref{fig:pot_scheme} shows, but because this approximation also assumes $d^* / d = \epsilon / \epsilon_{\text{sol}} \gg 1$ it is not valid for distances $r \ll d^*$.
At small distances $r \ll d$ the \RK potential instead reduces to the Coulomb potential with the bulk permittivity of the nanosheets, that in our units is simply

\begin{equation} \label{eq:r_ke_pot_short}
    V^{\text{RK}}(r \ll d) \simeq - \frac{d}{r} \equiv V^{\text{C}}(r)
\end{equation}

\noindent and which in S.I. units reads

\begin{equation}
    V^{\text{C}}(r) = - \frac{e^2}{4 \pi \epsilon} \frac{1}{r}
    \hspace{1cm}
    \text{ (in S.I. units)}
    .
\end{equation}

\begin{figure}[h!]
    \begin{center}
        \includegraphics[width=\linewidth]{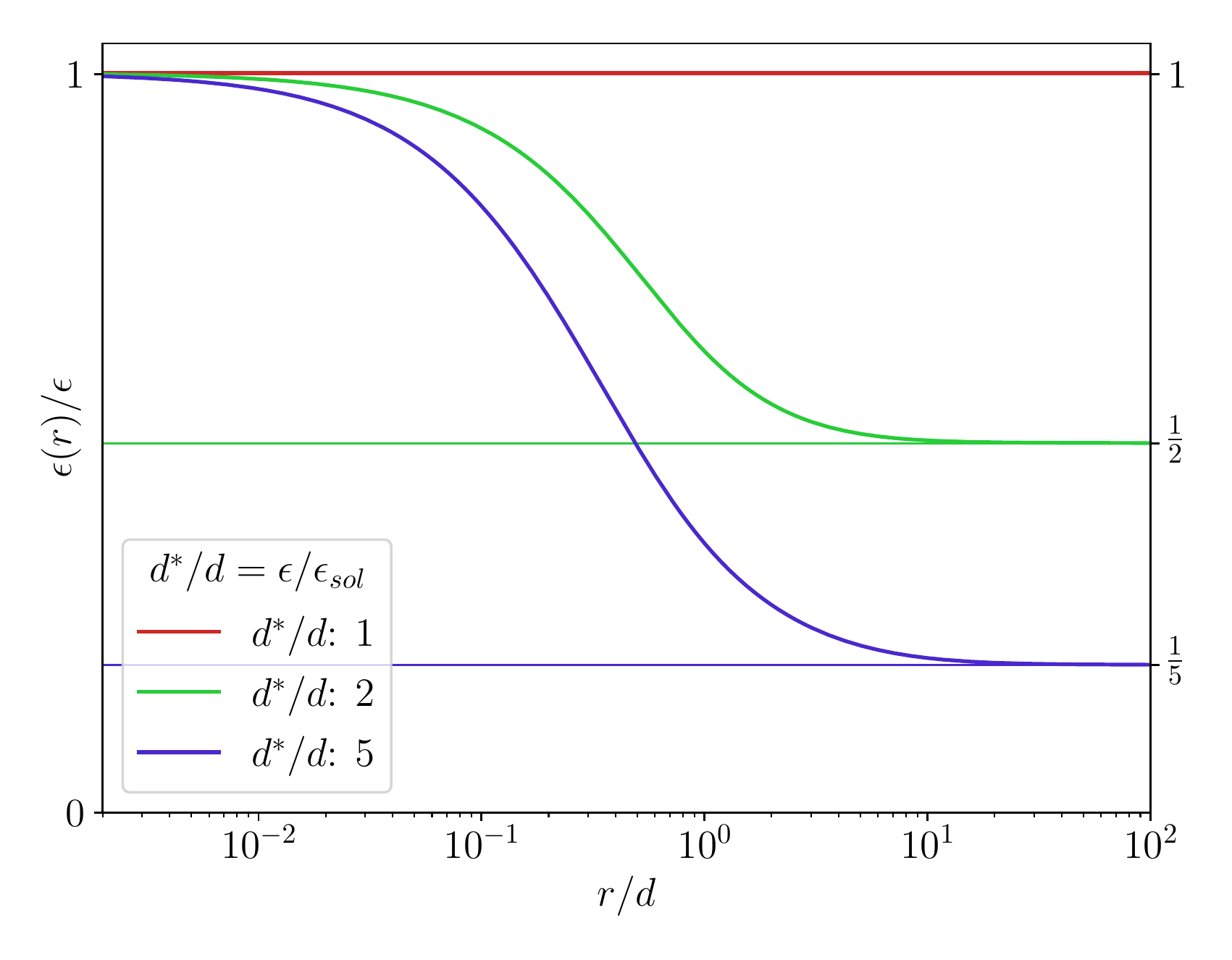}
        \caption{
            Real-space dependence of the dielectric function divided by the permittivity of the nanosheet $\epsilon(r) / \epsilon$, obtained from the \RK potential.
            At small distances $r / d \ll 1$ it saturates to one, which means that $\epsilon(r)$ reduces to the permittivity of the nanosheet.
            At large distances it saturates to $d / d^* = \epsilon_{\text{sol}} / \epsilon$, recovering the Coulomb potential of the solvent.
            The horizontal lines mark the corresponding saturation value.
        }
        \label{fig:pot_comparison_eps}
    \end{center}
\end{figure}

\begin{figure}[h!]
    \begin{center}
        \includegraphics[width=\linewidth]{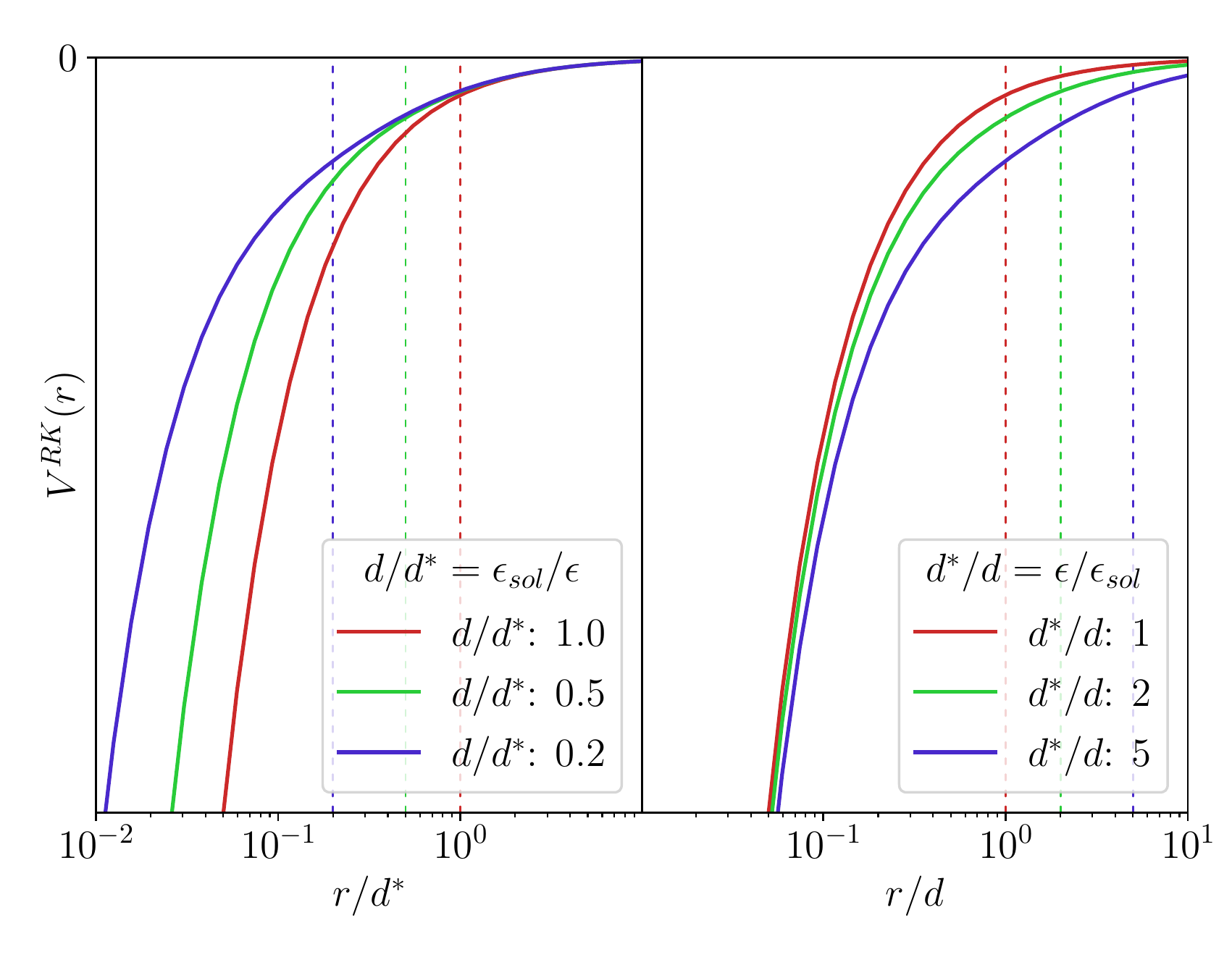}
        \caption{
            Real-space dependence of the \RK potential obtained by numerically Fourier transforming Eq. (\ref{eq:k_ke_pot}) for $z_1 = z_2 = 0$.
            The red lines (lowest one on the left, highest on the right) correspond to $d / d^* = 1$, in which case the \RK potential reduces to the Coulomb potential, and are the same on both plots.
            Vertical dashed lines mark the values $d / d^* = \epsilon_{\text{sol}} / \epsilon$ (left) and $d^* / d = \epsilon / \epsilon_{\text{sol}}$ (right).
            (Left) The \RK potential is shown in terms of $r / d^*$, thus the universal behavior is present for $r / d^* \gg 1$.
            (Right) The \RK potential is shown in terms of $r / d$, thus now the universal behavior is present for $r / d \ll 1$.
        }
        \label{fig:pot_comparison}
    \end{center}
\end{figure}

In order to present a more physical picture of the \RK potential, it may be interpreted as having a space-dependent dielectric function that connects the behavior at small and large distances, as

\begin{equation} \label{eq:r_ke_pot}
    V^{\text{RK}}(r) = - \frac{\epsilon}{\epsilon(r)} \frac{d}{r}
    ~~.
\end{equation}

\noindent where $\epsilon(r)$ is the dielectric function.
Equations (\ref{eq:r_ke_pot_long}), (\ref{eq:r_ke_pot_short}), and (\ref{eq:r_ke_pot}) suggest that $\epsilon(r \gg d) / \epsilon \rightarrow d / d^* = \epsilon_{\text{sol}} / \epsilon$ and $\epsilon(r \ll d) / \epsilon \rightarrow 1$, as is explicitly shown in Fig. \ref{fig:pot_comparison_eps}.
Physically, at small distances charges only feel the permittivity of the material as opposed to the surrounding solvent at large distances.
Note that the crossover in the dielectric function roughly takes place at $r \simeq d$, as expected physically.
Notice also that Eqs. (\ref{eq:r_ke_pot_long}) and (\ref{eq:r_ke_pot_short}) do not explicitly depend on the permittivities of the semiconductor material nor of the solvent, thus by using the variables $r / d$ and $r / d^*$ the behavior of the potential is universal for small and large distances respectively.
Figure \ref{fig:pot_scheme} shows the \RK potential and its approximations for each regime and Fig. \ref{fig:pot_comparison} shows such universal behavior for several values of $d^* / d = \epsilon / \epsilon_{\text{sol}}$.

Confining the electric field to two dimensions, as obtained by solving the purely two-dimensional Poisson equation, modifies the usual $1/r$ behavior of the Coulomb potential to a logarithmic behavior \cite{stoof2009}.
Since nanosheets are (quasi) two-dimensional one could expect that the interaction potential exhibits this logarithmic behavior at very small distances to a good approximation.
However, Figs. \ref{fig:pot_scheme} and \ref{fig:pot_comparison} clearly show that the \RK potential does not present such behavior, since electrostatics alone does not incorporate any two-dimensional confinement at small distances, and consequently it does not correctly describe the interaction between charges in this regime.

\section{\label{sec:eff_pot}Quantum confinement}

As a consequence of the \RK potential lacking a logarithmic behavior at small distances, it appears that the electrostatic approach alone does not incorporate the complete physics of the problem.
An important omission from this picture are the quantum-mechanical corrections to the potential, that is, the effect of the confined wave function of the interacting charges in the $z$ direction, which become significant when the inter-particle distance is of the order of the nanosheet thickness $d$ or less.

Consider a charge confined to the nanosheet in the direction of the axis perpendicular to the plane by an infinite well of size equal to the thickness $d$.
Solving for the $z$-component of the wave function in the Schr\"odinger equation with such a confining potential yields

\begin{equation}
    u_n(z) = \sqrt{\frac{2}{d}} \cos\left( \frac{n\pi z}{d}\right)
    ~~ \text{, for $n = 1, 3, 5, ...$}
    ~~ ,
\end{equation}

\noindent and

\begin{equation}
    u_n(z) = \sqrt{\frac{2}{d}} \sin\left( \frac{n\pi z}{d}\right)
    ~~ \text{, for $n = 2, 4, 6, ...$}
    ~~ ,
\end{equation}

\noindent with the corresponding energy

\begin{equation}
    E_n = \frac{\hbar^2}{2m} \left( \frac{n\pi}{d} \right)^2
    ~~ ,
\end{equation}

\noindent where $n$ is a positive integer.
If the energy difference between $E_1$ and $E_2$ is large enough compared to the interaction energy, i.e., $E_2 - E_1 \gg E^C(d^* / d)$, then excited states are not populated and only the ground state with wave function $u(z) \equiv u_1(z)$ alone determines the effect of quantum confinement.
In that case the thickness of the nanosheet has to satisfy

\begin{equation} \label{eq:d_max}
    d \ll \frac{4 \pi \epsilon_{\text{sol}}}{e^2} \frac{3 \pi^2 \hbar^2}{m} \equiv d_{\text{max}}
    ~~ .
\end{equation}

\noindent for the system to be (quasi) two dimensional.

The ground-state wave function determines the quantum-confined potential $V_{\text{qc}}^{\mathcal{V}}(k)$ as

\begin{equation} \label{eq:qc_gen_pot}
    V_{\text{qc}}^{\mathcal{V}}(k) = \hspace{-1.5mm}\int_{-\frac{d}{2}}^{\frac{d}{2}} \hspace{-2.5mm}\dif z_1 \int_{-\frac{d}{2}}^{\frac{d}{2}} \hspace{-2.5mm}\dif z_2 ~ u^2(z_1) u^2(z_2) V^\mathcal{V}(k, z_1, z_2)
    ~~ .
\end{equation}

\noindent Note that $V^\mathcal{V}(k, z_1, z_2)$ corresponds to the in-plane Fourier transform of a three-dimensional potential, that is

\begin{equation}
    V^\mathcal{V}(k, z_1, z_2) = \int \dif^2 \vec{r} ~ V^\mathcal{V}(r, z_1, z_2) e^{i \vec{k} \cdot \vec{r}}
    ~~.
\end{equation}

\noindent Here $\mathcal{V} = \text{C},~\text{RK}$ denotes either the Coulomb potential or the full \RK potential, respectively.

\subsection{\label{sec:eff_pot_cou}Coulomb potential}

Let us first apply this procedure to the Coulomb potential for the purpose of better understanding quantum confinement by itself, with the added advantage that our findings carry over to the \RK potential case as Sec. \ref{sec:eff_pot_rk} presents.
Using the Coulomb potential means that the nanosheet and the solvent have the same permittivity, i.e., $\epsilon_1 = \epsilon_2 = \epsilon$, which is not experimentally realizable, as the dielectric constant of a semiconductor typically obeys $\epsilon \gg \epsilon_{\text{sol}}$, that is, $d^* \gg d$, but sheds light on the mechanism that introduces the short-distance logarithmic behavior.
Note that in this case there is only one length scale, the thickness $d$, which is equivalently expressed as $d^* = d$.
Introducing the in-plane Fourier transform of the Coulomb potential

\begin{equation}
    V^{\text{C}}(k, z_1, z_2) = - \frac{2\pi d^2}{kd} ~ e^{-k|z_1 - z_2|}
\end{equation}

\noindent into Eq. (\ref{eq:qc_gen_pot}) yields

\begin{align}
    \label{eq:k_qccou_pot}
    \begin{split}
        V_{\text{qc}}^{\text{C}}(k) &= - \frac{2 \pi d^2}{kd} ~ \frac{3 (kd)^3 + 20 \pi^2}{ \left[(kd)^2 + 4 \pi^2\right]^2} \\
         &\phantom{=} ~- \frac{2 \pi d^2}{kd} ~ \frac{32 \pi^3}{(kd)^2} \frac{kd - 1 + e^{-kd}}{\left[(kd)^2 + 4 \pi^2\right]^2} \\
         &\equiv V_{<}(k) + V_{>}^{\text{C}}(k) ~~.
    \end{split}
\end{align}

\noindent The potential is separated into $V_{<}(k)$ and $V_{>}^{\text{C}}(k)$, each contribution dominating in the regimes $r \ll d$ and $r \gg d$ respectively.
Notice that the Fourier transform of any of the terms contained in $V_{>}^{\text{C}}(k)$ diverges if integrated separately from the others.

In the large-distance regime $r \gg d$ the quantum-confined Coulomb potential reduces to the Coulomb potential, the reason being that quantum confinement is significant only at small distances.
The long-distance behavior is obtained by expanding in powers of $kd$ and keeping the lowest-order term, yielding

\begin{equation} \label{eq:k_qccou_pot_long}
    V_{\text{qc}}^{\text{C}}(kd \ll 1) \simeq - \frac{2 \pi d^2}{k d} = V^{\text{C}}(k)
    ~~ .
\end{equation}

\noindent In the small-distance regime $r \ll d$ the contribution $V_{<}(r)$ dominates, up to a constant given by $V_{>}^{\text{C}}(r = 0)$.
An exact analytic expression for $V_{<}(r)$ is derived from Fourier transforming Eq. (\ref{eq:k_qccou_pot}), which results in

\begin{equation} \label{eq:r_qc_pot}
    V_{<}(r) = - 3 K_0\left(\frac{2 \pi r}{d}\right) + \frac{2 \pi r}{d} K_1\left(\frac{2 \pi r}{d}\right)
    ~~ ,
\end{equation}

\noindent where $K_\nu(z)$ is the modified Bessel function of the second kind.
As a consequence of the asymptotic behavior of $V_{<}(r)$ that tends to zero as $(r/d)^{\frac{1}{2}} e^{-r/d}$ for $r \rightarrow \infty$, its contribution is significant at small distances and exponentially suppressed otherwise.
The limit $K_\nu(r \rightarrow 0)$ reveals the short-distance logarithmic behavior as

\begin{equation}
    \label{eq:r_qc_pot_rlld}
    V_{<}(r \ll d) \simeq - 3 \left[ \ln\left(\frac{\pi r}{d}\right) + \gamma_E \right] - 1
    ~~ ,
\end{equation}

\noindent where $\gamma_E$ is the Euler-Mascheroni constant.
More physically, this logarithmic behavior is a direct consequence of the fact that the effective potential is obtained by averaging the bulk Coulomb potential over the $z$-position of the electron and the hole.

\begin{figure}[h!]
    \begin{center}
        \includegraphics[width=\linewidth]{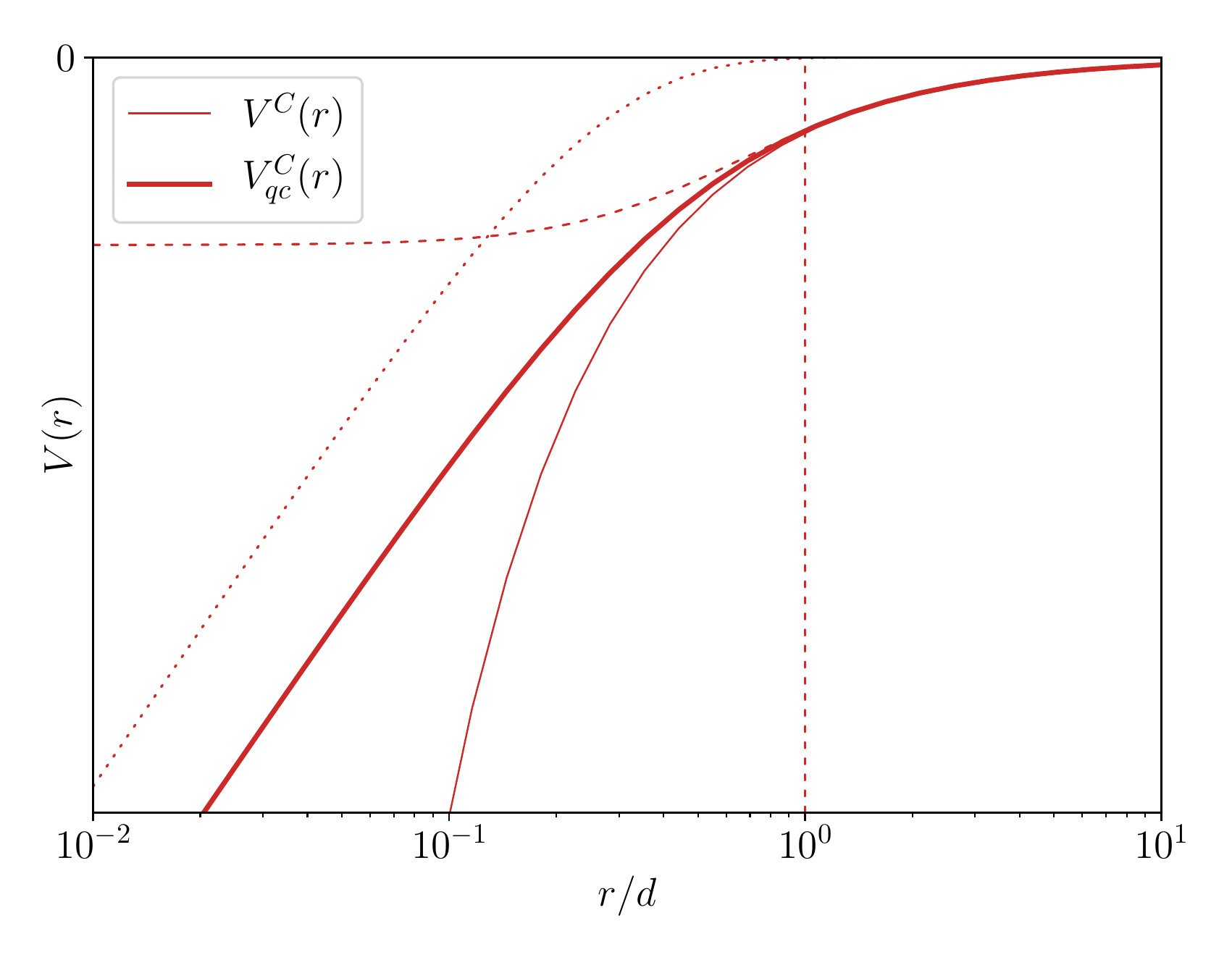}
        \caption{
            Real-space dependence of the quantum-confined Coulomb potential as a function of $r / d$, computed by Fourier transforming Eq. (\ref{eq:k_qccou_pot}).
            The two contributions $V_{<}(r)$ and $V_{>}^{\text{C}}(r)$ are represented by the dotted and dashed lines, respectively.
            The thin solid line corresponds to $V^{\text{C}}(r)$.
            The vertical dashed line correspond to the point $r / d = 1$.
        }
        \label{fig:r_qccou_pot}
    \end{center}
\end{figure}

Equation (\ref{eq:r_qc_pot_rlld}) shows some similarities between the quantum-confined Coulomb potential and the \RK potential at small distances, since both depend on the ratio $r / d$ alone, however $V_{\text{qc}}^{\text{C}}(r)$ incorporates the expected logarithmic behavior whereas the \RK potential does not.
In order to explicitly display the universality of the quantum-confined Coulomb potential, Fig. \ref{fig:r_qccou_pot} presents the result of numerically Fourier transforming Eq. (\ref{eq:k_qccou_pot}) to real space, as a function of $r / d$.

\subsection{\label{sec:eff_pot_rk}\RK potential}

With a better understanding of quantum confinement, let us explore the more realistic electrostatics situation described by the \RK potential.
Due to the disparate permittivities of nanosheet and solvent, the behavior of the \qcRK potential depends on the two distinct length scales $d$ and $d^*$.
Introducing the full \RK potential into Eq. (\ref{eq:qc_gen_pot}) yields

\begin{widetext}
    \begin{align}
        \label{eq:k_qcke_pot}
        \begin{split}
            V_{\text{qc}}^{\text{RK}}(k) &= V_{<}(k) - \frac{2 \pi d^2}{kd} ~ \frac{32 \pi^4}{(kd)^2} \\
            &\phantom{= V_{<}(k) }~ \times \frac{1}{\left[(kd)^2 + 4 \pi^2\right]^2} \left( kd - \frac{\sinh(kd + \eta_1)\sinh(\eta_2) + \sinh(kd + \eta_2)\sinh(\eta_1) - 2 \sinh(\eta_1)\sinh(\eta_2)}{\sinh(kd + \eta_1 + \eta_2)} \right) \\
            &\equiv V_{<}(k) + V_{>}^{\text{RK}}(k) ~~ ,
        \end{split}
    \end{align}
\end{widetext}

\noindent which is separated again into $V_{<}(k)$ and $V_{>}^{\text{RK}}(k)$, dominating in the small and large-distance regimes respectively.
Since $V_{<}(k)$ is identical as for the quantum-confined Coulomb potential, given in Eq. (\ref{eq:k_qccou_pot}), it follows that the logarithmic behavior is in fact introduced by quantum confinement regardless of the potential.

The tail of the \RK potential reemerges in the large-distance regime $r \gg d$ by expanding $V_{>}^{RK}(k)$ in powers of $kd$, while keeping only the lowest-order term, which results in

\begin{align}
    \begin{split}
        V_{\text{qc}}^{\text{RK}}(kd \ll 1) \simeq &- \frac{2 \pi d^2}{kd} ~ \frac{ \cosh(\eta_1)\cosh(\eta_2)}{\cosh(\eta_1 + \eta_2)} \\
        &\times \frac{2}{kd + \tanh(\eta_1 + \eta_2)}
        ~~ ,
    \end{split}
\end{align}

\noindent further simplifying to

\begin{equation} \label{eq:k_qcke_pot_long}
    V_{\text{qc}}^{\text{RK}}(kd \ll 1) \simeq - \frac{2 \pi (d^*)^2}{k d^*} \frac{1}{\frac{kd^*}{2} + 1} = V^{\hn}(k)
\end{equation}

\noindent in the case of $d^* / d = \epsilon / \epsilon_{\text{sol}} \gg 1$.
Equation (\ref{eq:k_qcke_pot_long}) shows that the \qcRK potential at large distances is approximated by the \HN potential that only depends on the length scale $d^*$.
In the small distance regime $r \ll d$ the logarithmic behavior of $V_{<}(r)$ takes over, while $V_{>}^{\text{RK}}(r)$ saturates to a constant at $r = 0$, similarly to the quantum-confined Coulomb potential.
Note, however, that in the truly two-dimensional limit $d \rightarrow 0$ the \qcRK potential reduces to the Coulomb potential of the solvent as given in Eq. (\ref{eq:r_ke_pot_long}).
The same also occurs for the \RK potential and the \HN potential.
Physically, this demonstrates that in the limit $d \rightarrow 0$ the effective permittivity is no longer determined by the semiconductor nanosheets, but only by the solvent.

\begin{figure}[h!]
    \begin{center}
        \includegraphics[width=\linewidth]{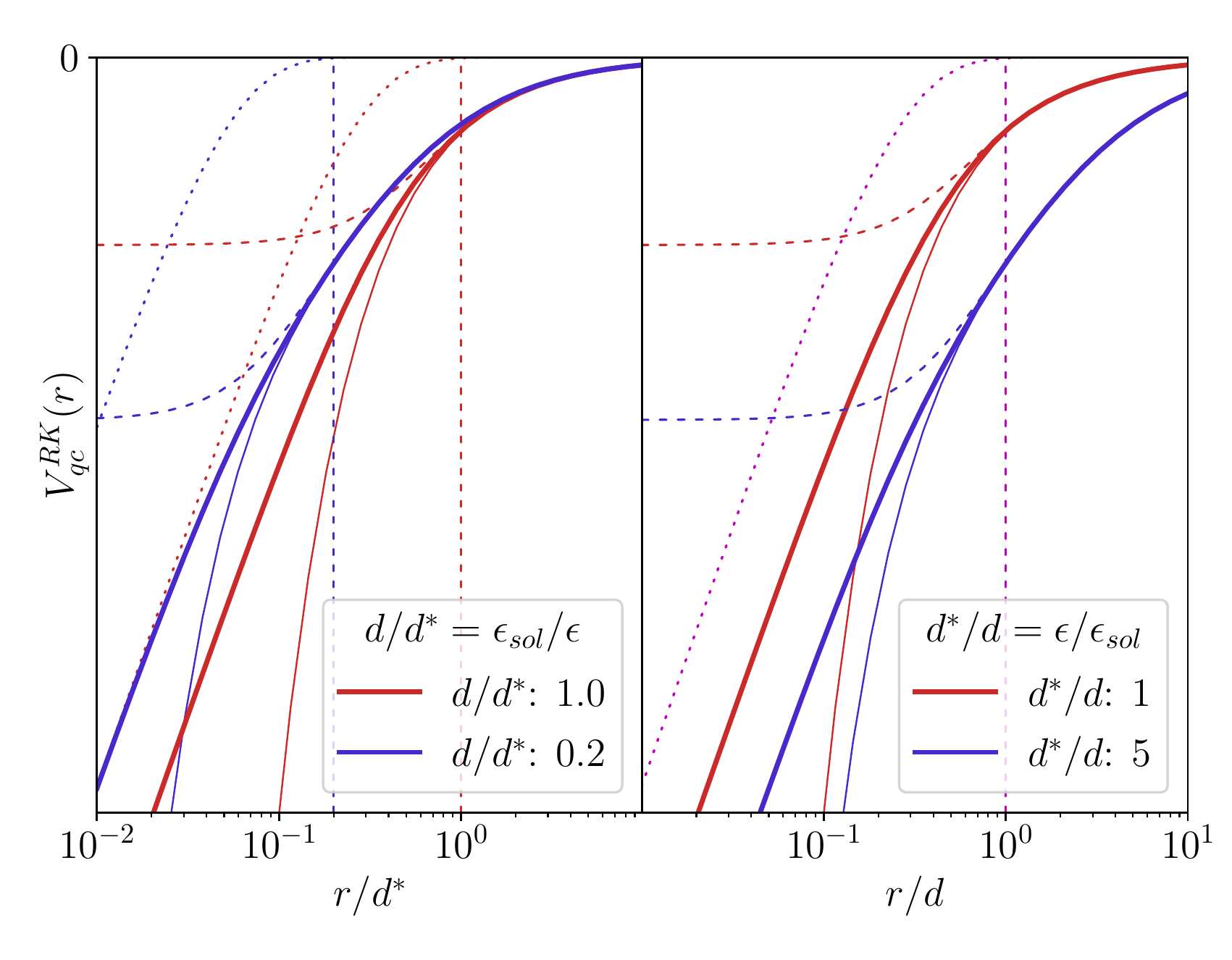}
        \caption{
            Real-space dependence of the \qcRK potential as a function of $r / d^*$ (left) and $r / d$ (right), computed by Fourier transforming Eq. (\ref{eq:k_qcke_pot}).
            The two contributions $V_{<}(r)$ and $V_{>}^{\text{RK}}(r)$ are represented by dotted and dashed lines, respectively.
            The thin solid lines correspond to $V^{\text{RK}}(r)$, for each value of $d / d^* = \epsilon_{\text{sol}} / \epsilon$ (left) and $d^* / d = \epsilon / \epsilon_{\text{sol}}$ (right).
            The red line (bottom one on the left, top one on the right, $d / d^* = 1$) is computed using the quantum-confined Coulomb potential, and thus is the same as that shown in Fig. \ref{fig:r_qccou_pot}.
            (Left) The vertical dashed lines correspond to the points $r / d^* = d / d^*$.
            (Right) The vertical dashed magenta line corresponds to $r / d = 1$.
        }
        \label{fig:r_qcke_pot}
    \end{center}
\end{figure}

Figure \ref{fig:r_qcke_pot} shows the real-space dependence of the \qcRK potential as a function of $r / d^*$ (on the left) and $r / d$ (on the right), analogously to Fig. \ref{fig:pot_comparison}, and presents the universal behavior in the regime $r / d^* \gg 1$ and $r / d \ll 1$ respectively.
Because $V_{<}(r)$ only depends on $r / d$, on the right side of Fig. \ref{fig:r_qcke_pot} it shows as only one line.

\begin{figure}[h!]
    \begin{center}
        \includegraphics[width=\linewidth]{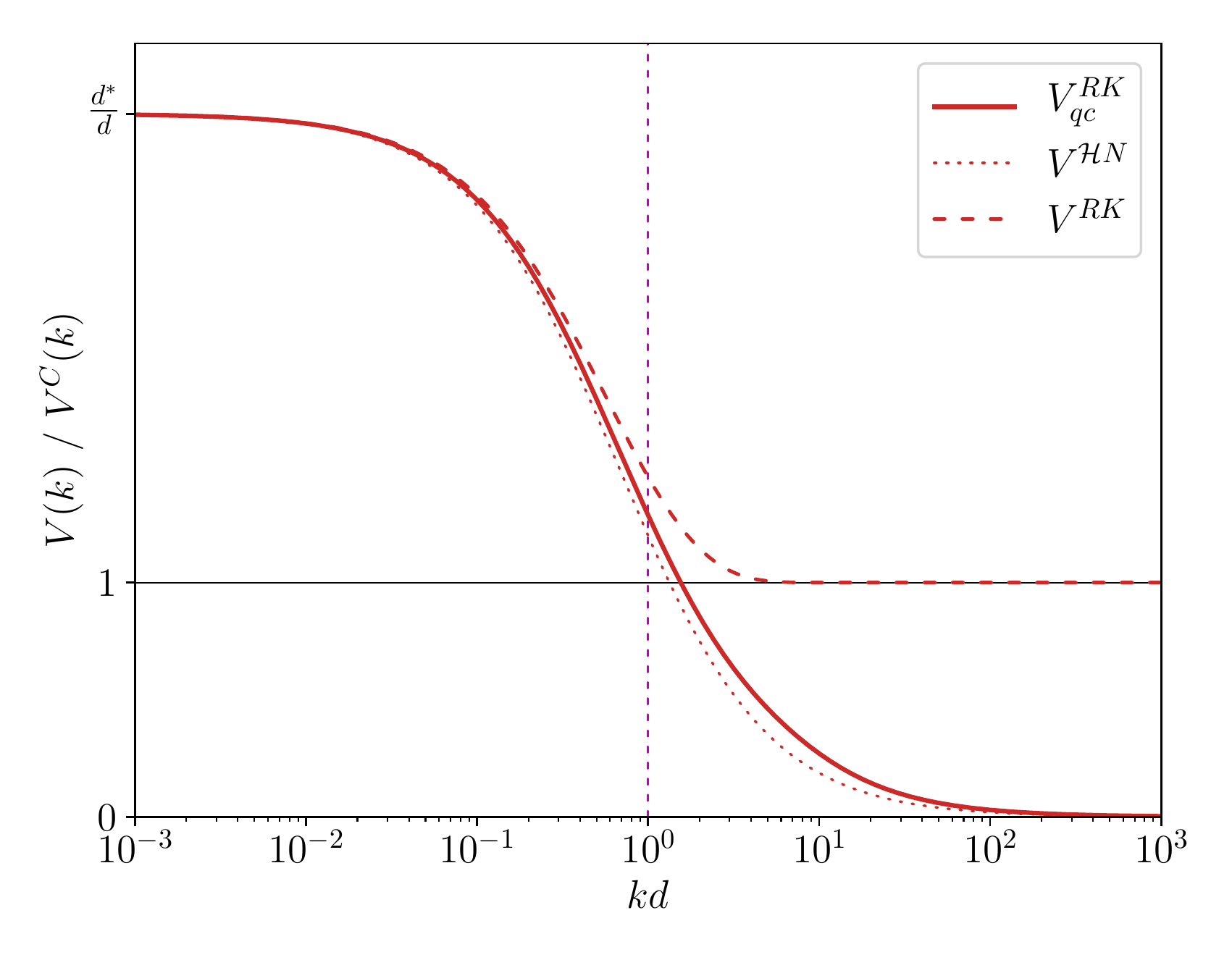}
        \caption{
            Momentum-space dependence of the \qcRK potential (solid), the \HN potential (dotted), and the \RK potential (dashed), divided by the Coulomb potential of the material, as a function of $kd$.
            The \qcRK potential is computed from Eq. (\ref{eq:k_qcke_pot}), the \HN potential from Eq. (\ref{eq:k_qcke_pot_long}), and the \RK potential from Eq. (\ref{eq:k_ke_pot}).
        }
        \label{fig:k_pot_comp}
    \end{center}
\end{figure}

Lastly, let us briefly consider the momentum-space dependence of each potential with the aim of understanding their differing small and large-distance behaviors from a new angle.
As an additional benefit introducing screening effects due to free charges is more straightforward in momentum space, as Ref. \cite{garciaflorez2019} extensively shows.
Figure \ref{fig:k_pot_comp} shows the momentum-space dependence of the \qcRK potential, the \HN potential, and the \RK potential, all divided by $V^{\text{C}}(k)$ as given in Eq. (\ref{eq:k_qccou_pot_long}).
In the regime $kd \ll 1$ every potential saturates to $d^* / d = \epsilon / \epsilon_{\text{sol}}$, which physically corresponds to recovering the Coulomb potential of the solvent at large distances.
On the opposite regime $kd \gg 1$ the tails differ significantly: for the \RK potential it tends to one, while for the \qcRK potential and the \HN potential it tends to zero.
In the case of the \RK potential the tail tends to one at large momenta, which means that the Coulomb potential of the material reappears at small distances.
Since the logarithmic behavior in real space appears due to Fourier transforming a term $\propto 1/k^2$, expressing the \RK potential in momentum space further shows that it does not incorporate a logarithmic tail at small distances.
For the \qcRK potential the tail does indeed tend towards zero as $1 / k^2$ due to $V_{<}(k)$, thus resulting in a logarithmic behavior at small distances.
Because the \HN potential is only but an approximation of the \RK potential for $kd \ll 1$, the behavior in the regime $kd \gg 1$ is in principle not valid as it is used outside of the region of applicability of this approximation, even though it resembles the \qcRK potential.

\section{\label{sec:excitons}Excitons}

Studying the small and large-distance behavior of the potentials does not by itself present the whole picture of the exciton wave functions as quantum mehcanics introduces another length scale into the problem due to the relative kinetic energy of an electron-hole pair.
Hence, we next compute the energy level of an exciton bound state that forms due to the \qcRK potential, the \HN potential, and the \RK potential with the intent of better understanding their differences and appropriately comparing our results with the literature.
Furthermore, we determine not only the ground-state exciton energy level but also that of the first several $s$-wave states, thus presenting a more complete analysis of our findings.

For the purpose of relating our results to an experimentally realizable case, we shift our focus towards CdSe nanoplatelets in hexane solvent at room temperature \cite{tomar2019,garciaflorez2019}.
As a consequence, we use for the electron and hole masses the values $m_e = 0.27~m_0$ and $m_h = 0.45~m_0$, where $m_0$ is the fundamental electron mass, corresponding to a thickness of 4.5 CdSe monolayers having $d=1.37$ nm, given as $n=4$ in Ref. \cite{benchamekh2014}.
The permittivity of the hexane solvent is $\epsilon_{\text{sol}} = 2 \epsilon_0$, where $\epsilon_0$ is the vacuum permittivity.
In an effort to present a more general discussion, the ratio of the material and the hexane permittivities $d^* / d = \epsilon / \epsilon_{\text{sol}}$ is treated as a free parameter, and the exciton energy levels are computed for several thicknesses.
Note that the effect that the thickness has on the effective masses is neglected here for simplicity without any impact on the qualitative behavior of our results.
Furthermore, any effects due to the finite lateral sizes of the nanoplatelets are also neglected.

Finding the exciton wavefunction $u_m(r)$ and accompanying energy $\mathcal{E}$ involves solving the radial part of the Schr\"odinger equation, given by

\begin{equation} \label{eq:schr_exc}
    \left( -\dert{}{r}{2} + \frac{m^2 - 1 / 4}{r^2} + \frac{d^*}{a_0} \frac{d^*}{d} \left( V(r d^*) - \mathcal{E} \right) \right) u_m(r) = 0
    ~~,
\end{equation}

\noindent with $m = 0$ corresponding to the $s$-wave solution of the exciton ground state on which we exclusively focus from now on.
Here we have scaled the equation by $E^C$, used $r \rightarrow r d^*$, and introduced the exciton Bohr radius $a_0$.
For a given material of permittivity $\epsilon$ the two-dimensional Bohr radius $a_0$ is \cite{yang1991}

\begin{equation} \label{eq:bohr_radius}
    a_0 \equiv \frac{2 \pi \epsilon}{e^2} \frac{\hbar^2}{m_r}
    ~~,
\end{equation}

\noindent where $m_r \equiv m_e m_h / (m_e + m_h)$ is the reduced mass of the exciton problem.
Hence in general, the two dimensionless parameters that the exciton energy turns out to depend on are $d^* / a_0$ and $d^* / d = \epsilon / \epsilon_{\text{sol}}$.
Notice that if the potential $V(r)$ is a function of the ratio $r / d^*$ alone, such as the Coulomb potential of the solvent or the \HN potential, then the dimensionless exciton energy $\mathcal{E}$ depends only on the product $(d^* / d) \cdot (d^* / a_0)$.
Lastly, note that the ratio $d^* / a_0$ does not depend on the dielectric constant of the material $\epsilon$, and thus it is equal to $d / a_0(\epsilon_{\text{sol}})$, i.e., the ratio of $d$ and the exciton Bohr radius in the Coulomb potential of the solvent.

\begin{figure}[h!]
    \begin{center}
        \includegraphics[width=\linewidth]{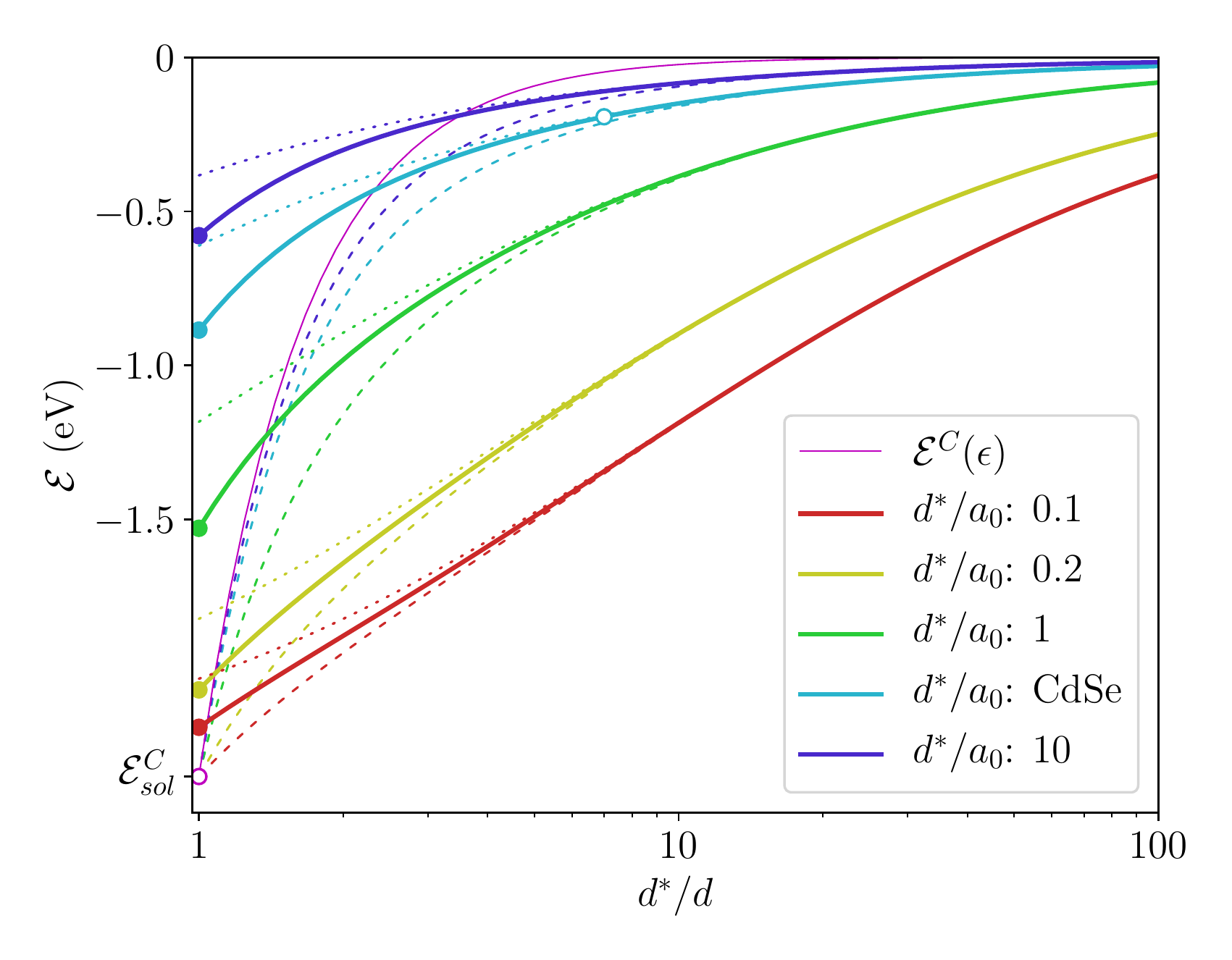}
        \caption{
            Exciton ground-state energy computed using the \qcRK potential $\mathcal{E}^{\text{RK}}_{\text{qc}}$ (solid), the \HN potential $\mathcal{E}^{\hn}$ (dotted), and the \RK potential $\mathcal{E}^{\text{RK}}$ (dashed).
            The filled dots mark the points for which $d^* / d = 1$, while the empty magenta dot marks $\mathcal{E}^{\text{RK}} = \mathcal{E}^{\text{C}}_{\text{sol}}$.
            The line labelled with ``CdSe'' corresponds to the thickness $d = 1.37$ nm ($d^* / a_0 \simeq 4.37$), and the empty light-blue dot corresponds to the ratio $d^* / d = \epsilon / \epsilon_{\text{sol}} \simeq 6.99$ that results in the exciton ground-state energy of $-193$ meV and exciton Bohr radius of $a_0 \simeq 2.19$ nm, representing the CdSe nanoplatelets considered in Refs. \cite{tomar2019,garciaflorez2019}.
            Lower values of $d^* / a_0$ result in more negative energies.
        }
        \label{fig:be_comp}
    \end{center}
\end{figure}

Let us analyze the exciton energies of CdSe in hexane in S.I. units that Fig. \ref{fig:be_comp} presents, obtained from computing the exciton ground-state energy as a function of the parameter $d^* / d = \epsilon / \epsilon_{\text{sol}}$, for several values of $d^* / a_0$.
Consider first the exciton energy obtained from the \RK potential $\mathcal{E}^{\text{RK}}$, represented by the dashed lines.
Because the \RK potential reduces to the Coulomb potential if $d^* / d = 1$, the exciton energy reduces to the hydrogen-like result

\begin{equation} \label{eq:be_cou_eps}
    \mathcal{E}^{\text{C}}(\epsilon) = - \frac{2 m_r e^4}{(4 \pi \epsilon \hbar)^2}
    ~~,
\end{equation}

\noindent that we define for the solvent as $\mathcal{E}^\text{C}_{\text{sol}} \equiv \mathcal{E}^{\text{C}}(\epsilon_{\text{sol}})$, and is represented by an empty magenta dot.
Furthermore, still focusing on the $d^* / d = 1$ case, notice that the \qcRK potential analogously reduces to the quantum-confined Coulomb potential, thus resulting in a smaller exciton energy $|\mathcal{E}^{\text{RK}}_{\text{qc}}| < \left|\mathcal{E}^\text{C}_{\text{sol}}\right|$ solely due to quantum confinement.
In order to satisfy the upper limit on the thickness set by Eq. (\ref{eq:d_max}) the parameter $d^* / a_0$ has to be small enough, which turns out to be $d \ll d_{\text{max}} \simeq 3.8$ nm for CdSe nanoplatelets.
This implies that we can consider only nanoplatelets with a thickess of at most 10 CdSe layers.
This condition is indeed satisfied for every curve shown in Fig. \ref{fig:be_comp}.
Regarding the \HN potential, it should be noticed that the resulting energy is not physically meaningful in the $d^* / d = 1$ case because it strictly speaking does not satisfy the assumption $d^* / d \gg 1$ used in the derivation of Eq. (\ref{eq:k_qcke_pot_long}).

Moving on to the regime $d^* / d > 1$, $\mathcal{E}^{\text{RK}}$ separates from $\mathcal{E}^{\text{C}}_{\text{sol}}$ as it becomes closer to zero.
Note that as the ratio $d^* / a_0$ grows, $\mathcal{E}^{\text{RK}}$ tends to $\mathcal{E}^{\text{C}}(\epsilon)$ due to the \RK potential reducing to the Coulomb potential of the material in the limit $d \rightarrow \infty$.
Furthermore, when $d^* / d \gg d^* / a_0$, i.e., $a_0 \gg d$, every one of the potentials considered in Fig. \ref{fig:be_comp} results in the same exciton energy, meaning that the small-distance behavior $r \ll d^*$ is no longer significant.

In the context of the dependence on the ratio $d^* / a_0$ we similarly identify two regimes.
For $d^* / a_0 \ll 1$ the electric field is mostly outside of the nanoplatelet, and consequently the exciton energy is close to that obtained using the Coulomb potential of the solvent $\mathcal{E}^{\text{C}}_{\text{sol}}$.
Only in the limit $d^* / d \gg 1$ the effect of the permittivity of the nanoplatelet significantly impacts the resulting energy.
In the case of $d^* / a_0 \gg 1$ the opposite is true: the permittivity of the nanoplatelet has a very significant effect on the exciton energy for any value of $d^* / d = \epsilon / \epsilon_{\text{sol}}$.

\begin{figure}[h!]
    \begin{center}
        \includegraphics[width=\linewidth]{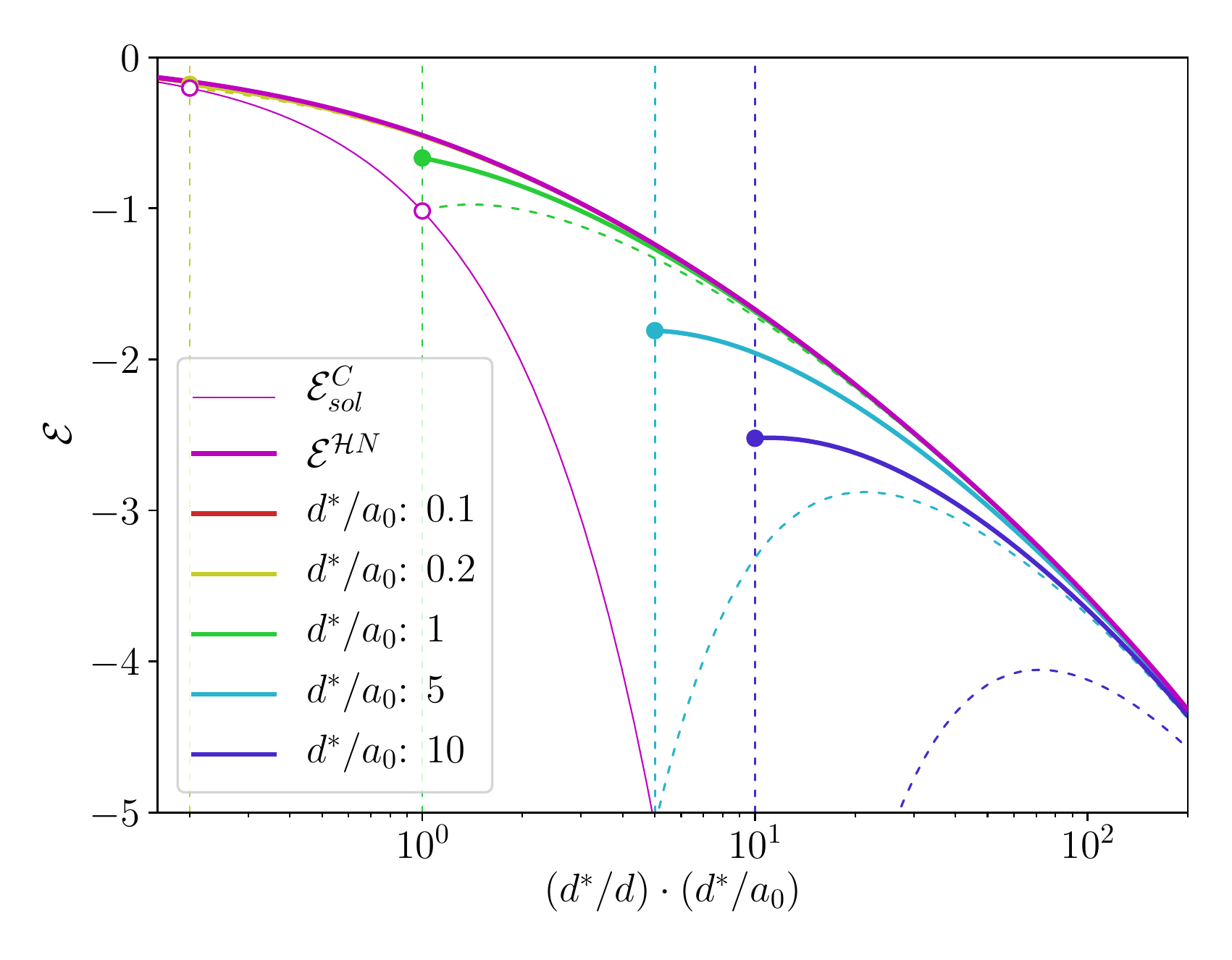}
        \caption{
            Exciton energy, scaled by $E^C \equiv e^2 / 4 \pi \epsilon d$, computed using the \qcRK potential (solid) and the \RK potential (dashed).
            Because the Coulomb potential of the solvent and the \HN potential only depend on $d^*$, the resulting energies $\mathcal{E}^{\text{C}}_{\text{sol}}$ and $\mathcal{E}^{\hn}$ (thin and thick magenta solid lines, respectively) do not depend on the ratio $d^* / a_0$.
            The filled dots mark the points for which $d^* / d = 1$, while the empty dots mark $\mathcal{E}^{\text{RK}} = \mathcal{E}^{\text{C}}_{\text{sol}}$.
            Higher values of $d^* / a_0$ result in more negative energies.
        }
        \label{fig:be_collapse}
    \end{center}
\end{figure}

Previously we discussed the universality of the \RK potential and \qcRK potential at small and large distances, made explicit when expressed in terms of the variables $r / d$ and $r / d^*$ respectively.
Even though this universality is also reproduced by the exciton energy, it is not immediately obvious in Fig. \ref{fig:be_comp}.
For the purpose of studying the universal behavior, the exciton energy is made dimensionless by $E^C \equiv e^2 / 4 \pi \epsilon d$ which results in a dimensionless exciton energy that we compute as a function of $(d^* / d) \cdot (d^* / a_0)$.
Figure \ref{fig:be_collapse} shows the exciton energies from Fig. \ref{fig:be_comp} altered by these transformations.
Despite the fact that in general $\mathcal{E}$ depends on $d^* / d = \epsilon / \epsilon_{\text{sol}}$ and $d^* / a_0$ separately, in the limit $d^* / d \gg 1$ the exciton energy turns out to be the same regardless of the potential used --- hence it depends on a single variable: the product of the two parameters as they appear in Eq. (\ref{eq:schr_exc}).
Consequently $\mathcal{E}_{\text{qc}}^{\text{RK}}$ shows a data collapse in the regime $d^* / d \gg d^* / a_0$, that is, the exciton Bohr radius is much larger than the thickness $a_0 \gg d$.

\begin{figure}[h!]
    \begin{center}
        \includegraphics[width=\linewidth]{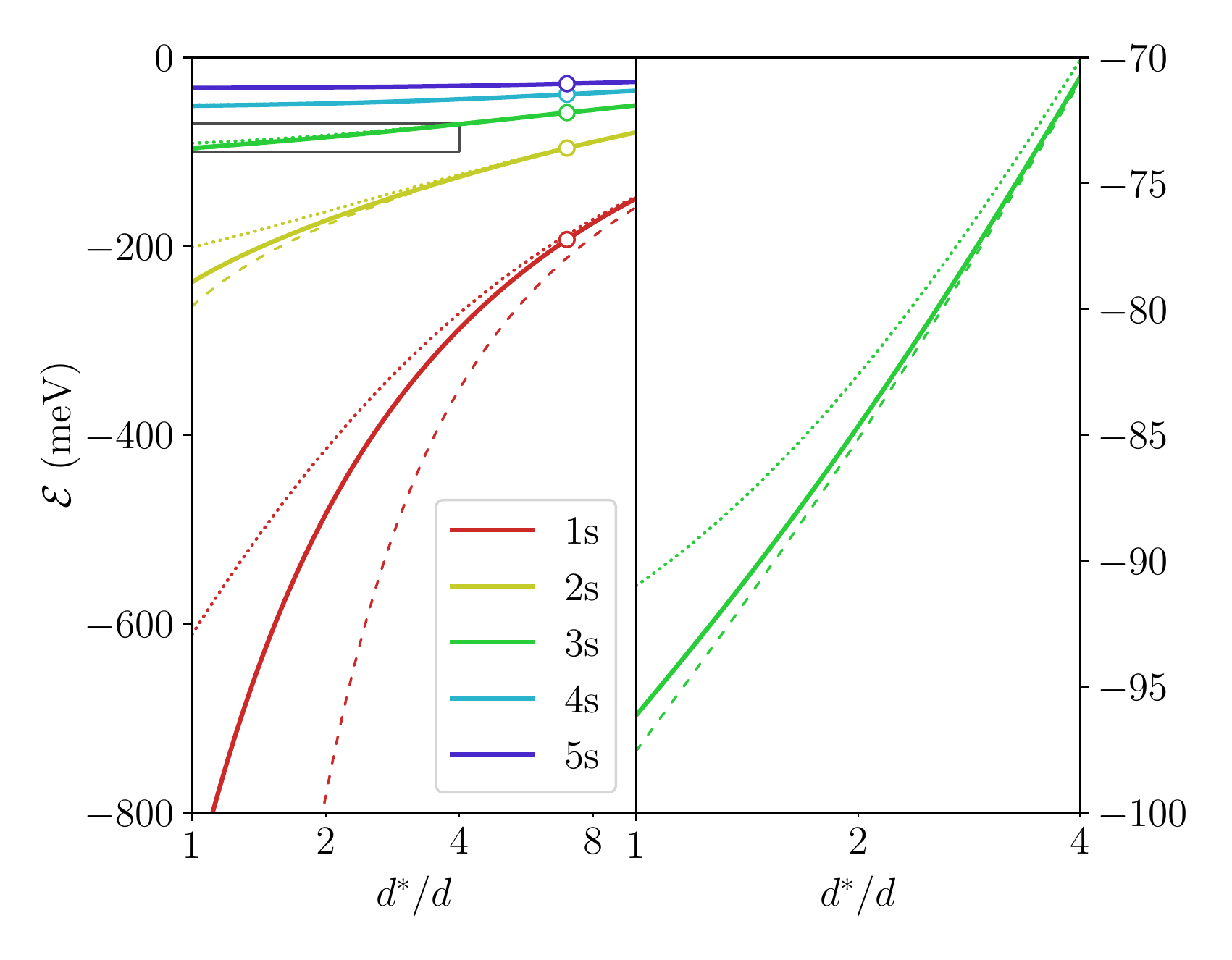}
        \caption{
            Energy of the first several $n$s states computed using the \qcRK potential (solid), the \HN potential (dotted), and the \RK potential (dashed).
            Computed for the thickness $d = 1.37$ nm ($d^* / a_0 \simeq 4.37$).
            On the left, the energies of the exciton ground state and the first four excited $n$s states.
            The empty dots mark the value of $d^* / d = \epsilon / \epsilon_{\text{sol}} \simeq 6.99$ that results in the exciton ground-state energy of $-193$ meV, representing the CdSe nanoplatelets considered in Refs. \cite{tomar2019,garciaflorez2019}.
            The energies of the excited states are $-96.1$ meV, $-58.6$ meV, $-39.2$ meV, and $-27.9$ meV.
            On the right, a zoom-in on the energy of the $3$s state, as an example, better showing the differences between potentials.
        }
        \label{fig:be_ns}
    \end{center}
\end{figure}

For a better comparison with other experimental results for CdSe in hexane solvent, Fig. \ref{fig:be_ns} shows not only the ground-state energy but also the energy of the first several excited $s$-wave states at the fixed ratio $d^* / a_0 \simeq 4.37$ corresponding to the experiment studied in Ref. \cite{tomar2019,garciaflorez2019}.
Furthermore, each energy is obtained from the \qcRK potential, the \HN potential, and the \RK potential.
Notice that the excited-states energies as a function of $d^* / d = \epsilon / \epsilon_{\text{sol}}$ generally follow the same trend as that of the ground state.
However, the right side of Fig. \ref{fig:be_ns} shows that using the \RK potential results in an energy closer to that obtained from the \qcRK potential than from the \HN potential.

\section{\label{sec:discussion}Summary and conclusion}

In an effort to elucidate the behavior of the \RK potential, we have presented a comprehensive discussion on the mechanism responsible for quantum confinement, alongside the two-dimensional short-distance logarithmic behavior of the interaction potential.
Section \ref{sec:electrostatics} began by presenting the actual behavior of the \RK potential at small distances, which led us to develop a method to incorporate quantum confinement into the potential.
Our approach is based on the charges being confined by an infinite well, but can easily be generalized to other situations that may be described by using a different confinement potential, in a similar direction as in Ref. \cite{latini2015}.
Of course, this would quantitatively affect our results for the quantum-confined potentials at small distances $r < d$, but not qualitatively.
With the \qcRK potential in hand, we analyzed its behavior in terms of the length scales $d$ and $d^* = \epsilon d / \epsilon_{\text{sol}}$ by comparing either the inter-particle distance $r$ or the two-dimensional exciton Bohr radius $a_0$ to these length scales.
Note that we have implicitly assumed that the exciton Bohr radius $a_0$, as well as every other length scale, is much larger than the lattice spacing of the material, that is, we only treat Wannier excitons, not Frenkel excitons.
In order to contrast our results with the literature, we provided an in-depth analysis of the exciton energy obtained using the \qcRK potential, in terms of the variables $d^* / d = \epsilon / \epsilon_{\text{sol}}$ and $d^* / a_0$.
In the future we aim to present a similar analysis for charges in the bulk and on the surface of a topological insulator.

\section*{Acknowledgments}

We would like to thank an anonymous referee of Ref. \cite{garciaflorez2019} for the helpful discussion that sparked the development of this paper.
This work is part of the research programme TOP-ECHO with project number 715.016.002, and is also supported by the D-ITP consortium.
Both are programmes of the Netherlands Organisation for Scientific Research (NWO) that is funded by the Dutch Ministry of Education, Culture and Science (OCW).


\appendix


%

\end{document}